# Recent Developments in Monte Carlo Simulations of First-Order Phase Transitions[*]

Wolfhard Janke

Institut für Physik, Johannes Gutenberg-Universität
D-55099 Mainz, Germany

**Abstract.** In the past few years considerable progress has been made in Monte Carlo simulations of first-order phase transitions and in the analysis of the resulting finite-size data. In this paper special emphasis will be placed on multicanonical simulations using multigrid update techniques, on numerical estimates of interface tensions, and on accurate methods for determining the transition point and latent heat.

## 1 Introduction

First-order phase transitions play an important role in many fields of physics [1–3]. Well-known examples are field-driven transitions in magnets, crystal melting, the nematic-isotropic transition in liquid crystals or, at much higher energy scales, the deconfining transition in hot quark-gluon matter and the various transitions in the evolution of the early universe [3].

An important property of first-order phase transitions is phase coexistence. For field-driven transitions as, e.g., in the Ising model at low temperatures or the $\phi^4$ theory discussed below, this is reflected in the canonical ensemble by a highly double-peaked probability distribution $P_{\rm can}(m)$ of the magnetization $m$. To sample in Monte Carlo (MC) simulations the two peaks with the correct relative weight the system has to pass many times through mixed phase configurations. For finite periodic systems of size $L^d$, such configurations contain (at least) two interfaces which contribute an additional term $2\sigma L^{d-1}$ to the free energy, where $\sigma$ is the interface tension. Compared to the peak maxima they are hence suppressed by an additional Boltzmann factor $\propto \exp\{-2\sigma L^{d-1}\}$ which implies an exponential divergence of the autocorrelation time with system size, $\tau \propto \exp\{2\sigma L^{d-1}\}$. This property is sometimes called supercritical slowing down. The same arguments apply to temperature-driven first-order phase transitions as, e.g. in Potts models, where $m$ has to be replaced by the energy $e$ and $P_{\rm can}(m)$ by $P_{\rm can}(e)$.

Much effort has been spent in the past few years to develop efficient methods for numerical studies of this important class of phase transitions. Both, improved update schemes for data generation and refined techniques for data analysis have been studied. To overcome the slowing down problem, a so-called multicanonical method has been proposed which is basically a reweighting approach and can, in principle, be combined with any legitimate update algorithm. To date most applications employed the local heat-bath or Metropolis algorithms. These studies showed that (at least for all practical purposes) the exponential slowing down is indeed reduced to a much weaker power-law divergence with increasing system size. Since the remaining slowing down problem is, however, still severe, several studies have tried to combine the multicanonical approach with other update algorithms which are known to be much more efficient in the case of continuous phase transitions. In Sec. 2, after a brief review of the multicanonical method, special emphasis will be laid on a recently proposed combination with multigrid update algorithms which was shown to give a further real-time improvement of about one order of magnitude for a two-dimensional $\phi^4$ lattice model.

The uniform accuracy of the probability distribution in multicanonical simulations led to reinvestigations of a long-known histogram technique to determine the interface tension between the coexisting phases at the transition point. The 2D $q$-state Potts model, the 2D and 3D Ising model and the 2D $\phi^4$ lattice model as well as the (3+1)D SU(3) lattice gauge theory have been studied so far. In Sec. 3.1 a summary is given of numerical results for the interface tension of 2D Potts models which can be compared with a recently derived analytical formula. This formula relies on an exact expression for the correlation length in the disordered phase at the transition point. Some direct numerical tests of this expression are presented in Sec. 3.2.

Parallel to the algorithmic developments many exact results concerning the finite-size scaling behavior at (strong) first-order phase transitions could be proven. This formulation suggested refined methods to estimate the transition point and latent heat from finite-size data, which are discussed in Sec. 3.3. The important point is that these estimates are exponentially close to the infinite-volume limit, i.e., they do not show the typical power-law corrections $\propto 1/L^d$ of the traditional observables and are hence of improved accuracy.

Finally, Sec. 4 contains a few concluding remarks.

## 2 Improved Generation of Monte Carlo Data

### 2.1 Multicanonical reweighting

The slowing down problem at first-order phase transitions is directly related to the double-peak shape of the canonical probability distribution $P_{\rm can}(m)$ or $P_{\rm can}(e)$. In multicanonical simulations [4, 5] of field-driven transitions this problem is avoided by simulating an auxiliary distribution $P_{\rm muca}(m) = P_{\rm can}(m) \times \exp(-f(m))$, where the reweighting factor $\exp(-f(m)) \equiv w(m)^{-1}$ is adjusted iteratively until $P_{\rm muca}(m) = const.$ between the two peaks [6–8]. This gives the mixed phase configurations the same statistical weight as the pure phases. Precisely the same idea applies to temperature-driven transitions with $m$ replaced by $e$. Canonical expectation values $\langle\mathcal{O}\rangle_{\rm can}$ of any observable $\mathcal{O}$ can be recovered by the reweighting formula $\langle\mathcal{O}\rangle_{\rm can} = \langle w\mathcal{O}\rangle_{\rm muca}/\langle w\rangle_{\rm muca}$, where $\langle\ldots\rangle_{\rm muca}$ denote expectation values with respect to the multicanonical distribution $P_{\rm muca}$.

Using local algorithms to update the multicanonical distribution it was demonstrated for various models [9–16] that the exponential slowing down is indeed reduced to a power-law behavior $\tau \propto V^\alpha = L^{d\alpha}$ with $\alpha \approx 1 - 1.5$, as one would expect from a simple random walk argument. While this is clearly an important step forward, the remaining slowing down problem is still severe. In fact, it is even worse than in simulations of critical phenomena where standard local MC algorithms yield $\tau \propto L^z$ with $z \approx 2$ [17, 18]. Here, however, several improved (mostly non-local) update algorithms (overrelaxation, cluster, multigrid, ...) are available which greatly reduce or even completely eliminate the slowing down problem [19–21].

---

[*] To appear in *Computer Simulations in Condensed Matter Physics VII*, eds. D.P. Landau, K.K. Mon, and H.B. Schüttler (Springer Verlag, Heidelberg, Berlin, 1994).

In order to further improve the performance of Monte Carlo simulations of first-order phase transitions it is therefore quite natural to study combinations of multicanonical reweighting with these improved update algorithms. Overrelaxation techniques have been used in the context of SU(3) lattice gauge theory [16], but due to the complexity of this system no systematic investigation of the performance has been reported. Cluster updates are difficult to implement in a straightforward way since, due to the reweighting factor $f(m)$ or $f(e)$, the multicanonical Hamiltonian is implicitly non-local. The more involved solution [11] proposed for Potts models yields a reduced exponent $\alpha$ and should thus be favorable for large systems. For moderate lattice sizes, however, it is again not clear how much is gained in real computer time. Multigrid update techniques finally can be quite efficient for problems with continuous fields as will be discussed in the next subsections.

## 2.2 Multigrid Monte Carlo Update Schemes

The general strategy of multigrid MC update schemes [22–24] is to perform collective moves on successively longer length scales in a systematic order as extensively discussed in the context of partial differential equations [25]. Both the type of collective moves and the sequence of length scales are parameters of multigrid schemes which can be defined in two equivalent ways: in a *unigrid* formulation where the update scheme is described in terms of the field variables on the original (fine) lattice, or in a *recursive multigrid* formulation where successively coarser lattices and associated Hamiltonians are introduced [19]. While the unigrid formulation is very transparent and easy to program, the more involved recursive multigrid formulation is numerically much more efficient.

In a unigrid formulation the collective update proposals are usually taken as square excitations, that is all fields in successively larger blocks of size $1, 2^d, 4^d, \ldots, V = 2^{nd}$ are moved by the same amount $\phi_b$. Another possibility are excitations of the form of pyramids where $\phi_b$ is the amplitude of the central displacement. We then accept or reject the proposed collective move in accordance with the usual Metropolis formula. In a recursive multigrid formulation the block amplitudes $\phi_b$ are identified with field variables $\phi_i^{(k)}$ on coarse lattices $\Xi^{(k)}$ (with $2^{kd}$ sites), and the shape of the excitations is controlled by an operator $\mathcal{P}$ which interpolates the coarse grid fields back to the next finer grid $\Xi^{(k+1)}$. The square excitations correspond to a piecewise constant interpolation ($\mathcal{P}(\phi_1^{(k)}, \phi_2^{(k)}, \ldots) = (\phi_1^{(k)}, \phi_1^{(k)}, \phi_2^{(k)}, \phi_2^{(k)}, \ldots)$), and the pyramids correspond to a piecewise linear interpolation. To update the fields $\phi_i^{(k)}$ one defines a coarse grid Hamiltonian recursively as $H^{(k)}(\phi_i^{(k)}) = H^{(k+1)}(\phi_i^{(k+1)} + \mathcal{P}\phi_i^{(k)})$, $H^{(n)} = H$, and initializes the $\phi_i^{(k)}$ to zero. An update sweep over the coarse grid $\Xi^{(k)}$ then corresponds to updating the amplitudes $\phi_b$ of blocks of size $2^{(n-k)d}$ in the unigrid formulation. The multigrid formulation goes actually one step further and also defines the sequence of coarse grids or length scales in a recursive way. More precisely the update on level $k$ thus consist of a) $m_1$ presweeps using any valid update algorithm with Hamiltonian $H^{(k)}$, b) calculating the Hamiltonian for the next coarser grid and initializing the field variables $\phi_i^{(k-1)}$ to zero. One then c) updates the field variables $\phi_i^{(k-1)}$ by applying the multigrid update $\gamma_{k-1}$ times, d) interpolates the variables of level $k-1$ back to the finer grid, and e) performs another $m_2$ postsweeps on level $k$. On the coarsest grid $\Xi^{(0)}$ only steps a) and e) can be performed and the recursion stops. At step c) the sequence of coarse grids is defined by the cycle parameters $\gamma_k$. For $\gamma_k \equiv \gamma = 1$ (V-cycle) every grid is given the same weight, while for $\gamma_k \equiv \gamma = 2$ (W-cycle) the coarser grids are updated more frequently. The cycle names derive from their graphical representations which very much resemble the letters V and W; see Fig. 1.

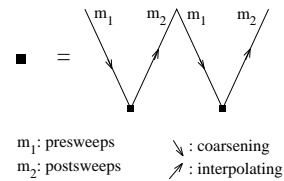
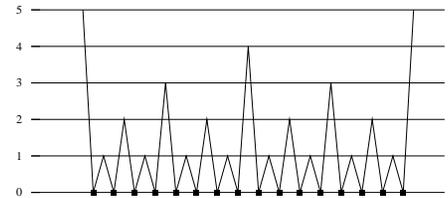

Figure 1: The recursive definition of the W-cycle and the expanded W-cycle for $L = 32$.

For the Gaussian model it can be proven [19] that critical slowing down is completely eliminated a) for piecewise linear interpolation (or higher) and any cycle or b) for piecewise constant interpolation and the W-cycle (or higher). For a V-cycle with piecewise constant interpolation this has not yet been proven and, in fact, numerical evidence indicates that in this case $z = 1$ [26, 27].

An important feature of the recursive multigrid formulation is the fact that the computational labor $W$ per cycle is proportional to the work $W_0$ per sweep of the original lattice, $W \approx W_0 (1 - \gamma/2^d)^{-1}$ ($\gamma < 2^d$), independent of the lattice size. In the unigrid formulation, on the other hand, the estimates are $W \approx W_0 \log_2 L$ for the V-cycle and $W \approx W_0 L^{\log_2 \gamma}$ for $\gamma \geq 2$, i.e, $W \approx W_0 L$ for a W-cycle.

## 2.3 Multicanonical Multigrid Monte Carlo Method

When adapting multigrid schemes to multicanonical simulations one only has to make sure that also the reweighting factor $f(m)$ or $f(e)$ is treated properly [27–29]. Let us first consider the case of a field-driven transition in the unigrid viewpoint. Using piecewise constant interpolation, a proposed move $\phi_b$ for a block of size $2^{kd}$ would change the magnetization by $\Delta m = 2^{kd} \phi_b / V$. The decision of acceptance is thus simply to be based on the value of $\Delta E_{\text{muca}} = \Delta E_{\text{can}} + f(m + 2^{kd} \phi_b V) - f(m)$. In the recursive multigrid implementation the multicanonical modification is very similar since the coarse grid magnetization can be computed in the standard way. For a temperature-driven transition $\Delta m$ would have to be replaced by $\Delta E_{\text{can}}/V$, which is also straightforward to implement in both the unigrid and multigrid formulation. For a discussion of more general situations, see Ref.[29].

## 2.4 Application to the 2D $\phi^4$ Model

As an application of the multicanonical multigrid algorithm we considered in Ref. [29] the scalar $\phi^4$ lattice field theory defined by the partition function

$$Z = \prod_{i=1}^{V} \left[ \int_{-\infty}^{\infty} d\phi_i \right] \exp\left( -\sum_{i=1}^{V} \left( \frac{1}{2}(\vec{\nabla}\phi_i)^2 - \frac{\mu^2}{2}\phi_i^2 + g\phi_i^4 \right) \right), \quad (1)$$

with $g = 0.25$ and $\mu^2 > \mu_c^2(g) > 0$. Here $\mu_c^2(g)$ is a line of critical points separating the disordered ($\mu^2 < \mu_c^2(g)$) and ordered ($\mu^2 > \mu_c^2(g)$) phase. For $d \geq 2$ and $\mu^2 > \mu_c^2(g)$, the reflection symmetry is spontaneously broken in the infinite volume limit $L \to \infty$, and the average magnetization $\langle m \rangle = \langle (1/V) \sum_{i=1}^{V} \phi_i \rangle$ acquires a nonvanishing expectation value. Consequently, if a term $h \sum_i \phi_i$ is added to the energy, the system exhibits first-order phase transitions driven by the field $h$.

We have studied the field-driven first-order phase transition between the two ordered phases in $d = 2$ dimensions at $g = 0.25$ for $\mu^2 = 1.30, 1.35$, and $1.40$. These points are sufficiently far away from the critical point at $\mu_c^2 = 1.265(5)$ [30] to display the typical behavior already on quite small lattices. A sensitive measure of the strength of the transition is the order-order interface tension $\sigma_{oo}$ between the $+$ and $-$ phase, which turns out [29] to be $\sigma_{oo} = 0.03443(47), 0.09785(60)$, and $0.16577(73)$ [31].

To evaluate the performance of the multicanonical multigrid algorithm, we recorded the time series for several observables $\mathcal{O}$ and studied their autocorrelation functions $A(j) = \rho(j)/\rho(0)$, where $\rho(j) = \langle \mathcal{O}_i \mathcal{O}_{i+j} \rangle - \langle \mathcal{O}_i \rangle^2$. The exponential autocorrelation time $\tau^{\exp}$ follows from its asymptotic decay for large $j$, $A(j) \propto \exp(-j/\tau^{\exp})$, and the integrated autocorrelation time $\tau^{\text{int}}$ is obtained in the large $k$ limit of $\tau(k) = 1/2 + \sum_{j=1}^{k} A(j)$. Here we shall concentrate only on the magnetization $m$ which reflects most directly the (pseudo) dynamics of the transitions between the two ordered phases. In general we are interested in the variance $\epsilon^2$ of the (weakly biased) estimator $\hat{\mathcal{O}} = \sum_{i=1}^{N_m} w(m_i) \mathcal{O}_i / \sum_{i=1}^{N_m} w(m_i) \equiv \overline{w_i \mathcal{O}_i}/\overline{w_i}$ for a canonical expectation value $\langle \mathcal{O} \rangle_{\text{can}}$. To facilitate a direct comparison with canonical simulations, we *define* for multicanonical simulations an effective autocorrelation time $\tau^{\text{eff}}$ by the standard error formula for $N_m$ correlated measurements,

$$\epsilon^2 = \sigma_{\text{can}}^2 2\tau^{\text{eff}}/N_m, \quad (2)$$

where $\sigma_{\text{can}}^2 = \langle \mathcal{O}_i^2 \rangle_{\text{can}} - \langle \mathcal{O}_i \rangle_{\text{can}}^2$ is the variance of single measurements in the canonical distribution. The squared error $\epsilon^2$ can be estimated either directly by jack-knife blocking procedures [32], or by applying standard error propagation to $\hat{\mathcal{O}} = \overline{w_i \mathcal{O}_i}/\overline{w_i}$, which involves the (multicanonical) variances and covariances of $w_i \mathcal{O}_i$ and $w_i$, and the three associated integrated autocorrelation times $\tau^{\text{int}}_{\mathcal{O};\mathcal{O}} \equiv \tau^{\text{int}}_{\mathcal{O}}$, $\tau^{\text{int}}_{w\mathcal{O};w\mathcal{O}} \equiv \tau^{\text{int}}_{w\mathcal{O}}$, and $\tau^{\text{int}}_{w\mathcal{O};\mathcal{O}} = \tau^{\text{int}}_{\mathcal{O};w\mathcal{O}}$ [28, 29].

Some results for $\mu^2 = 1.30$ from runs with at least $20\,000\,\tau^{\text{int}}_{wm}$ Metropolis sweeps or W-cycles (using piecewise constant interpolation without postsweeps) are collected in Table 1. We see that $\tau^{\exp}_m$ and $\tau^{\text{int}}_m$ agree well with each other, indicating that the corresponding autocorrelation function $A(j)$ can be approximated by a single exponential. While the $\tau^{\exp}_{wm}$ agree with $\tau^{\exp}_m$ within error bars, the integrated autocorrelation times $\tau^{\text{int}}_{wm}$ are significantly smaller. This implies that for this observable $A(j)$ is composed of many different modes. We further observe that the difference between $\tau^{\text{int}}_{wm}$ and $\tau^{\text{eff}}_m$ can be quite appreciable, reflecting the varying probability distribution shapes with increasing $L$. For comparison with previous work we have also included the diffusion times $\tau^{\text{flip}}_m$, which are defined as one quarter of the time it takes to travel from one (canonical) peak to the other and back.

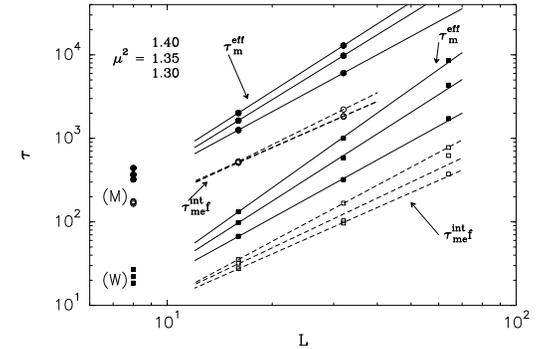

Figure 2: Effective autocorrelation times $\tau^{\text{eff}}_m$ for the model (1) in $d = 2$ with $g = 0.25$ as a function of lattice size $L$.

The log-log plot in Fig. 2 shows the behavior of $\tau^{\text{eff}}_m$ as a function of the lattice size $L$. Least-squares fits of $\tau^{\text{eff}}_m$ to a power law, $\tau^{\text{eff}}_m \propto V^\alpha$, yield for both update algorithms exponents of $\alpha \approx 1.2, 1.4$, and $1.5$ for $\mu^2 = 1.30, 1.35$ and $1.40$, i.e., the multigrid update does not improve the asymptotic behavior. The effective autocorrelation times of the W-cycle, however, are about 20 times smaller than those of the Metropolis algorithm. If we finally take into account that a W-cycle requires more elementary operations than a Metropolis sweep, we obtain with our implementations on a CRAY Y-MP a *real time* improvement factor of about 10.

Table 1: Autocorrelation times in units of sweeps resp. cycles for the Metropolis (M) and multigrid W-cycle (W) update in multicanonical simulations of the model (1) in $d = 2$ with $g = 0.25$ and $\mu^2 = 1.30$.

|  | $L = 8$ |  | $L = 16$ |  | $L = 32$ |  | $L = 64$ |
|---|---|---|---|---|---|---|---|
|  | M | W | M | W | M | W | W |
| $\tau^{\exp}_m$ | 212(12) | 11.30(32) | 668(23) | 37.2(2.0) | 3120(200) | 148(11) | 746(62) |
| $\tau^{\text{int}}_m$ | 204.4(4.0) | 10.88(12) | 690(11) | 34.69(76) | 2984(63) | 150.0(4.0) | 758(37) |
| $\tau^{\exp}_{wm}$ | 209(12) | 11.34(33) | 655(31) | 36.9(2.0) | 2880(190) | 146(13) | 600(120) |
| $\tau^{\text{int}}_{wm}$ | 171.1(3.4) | 9.82(11) | 509.8(8.9) | 27.58(59) | 1840(40) | 96.6(2.4) | 374(23) |
| $\tau^{\text{eff}}_m$ | 322.7(6.1) | 18.51(20) | 1258(21) | 67.4(1.3) | 6050(120) | 321.9(7.6) | 1724(86) |
| $\tau^{\text{flip}}_m$ | 463.5(6.4) | 30.82(25) | 1759(24) | 91.7(1.3) | 7780(140) | 428.2(8.9) | 1922(85) |

# 3 Refined Data Analysis

## 3.1 Interface Tension

A quantity of central importance for the kinetics of first-order phase transitions is the interface tension $\sigma$ between the coexisting phases [1, 2]. As discussed earlier, on finite periodic lattices of size $L^d$, this is reflected by a double-peak structure of the probability distribution for the energy or magnetization, with the minimum between the two peaks dominated by mixed phase configurations with two interfaces contributing an excess free energy of $2\sigma L^{d-1}$. This suggests [43] to extract the interface tension from the infinite volume limit of

$$2\sigma^{(L)} = \frac{1}{L^{d-1}} \ln(P_{\max}^{(L)}/P_{\min}^{(L)}). \quad (3)$$

For accurate results the system has to travel many times between the two peaks, which is a serious problem in canonical simulations of strong first-order transitions (large $\sigma$). The multicanonical sampling, on the other hand, is just designed for this purpose since it gives the same relative errors for $P_{\max}^{(L)}$ and $P_{\min}^{(L)}$ and thus optimizes the error on $\sigma^{(L)}$.

Using this so-called histogram method interface tensions have been estimated at the temperature driven first-order phase transitions of 2D $q$-state Potts models [9–12, 14, 51] and $N_t = 2$ SU(3) lattice gauge theory [10, 16], and at the field-driven transitions of the 2D and 3D Ising [15] and the 2D $\phi^4$ model [29]. For the 2D Ising model good agreement with the exact Onsager formula was obtained. For the 2D $q$-state Potts model with Hamiltonian [34]

$$H = -J \sum_{\langle ij \rangle} \delta_{s_i s_j}; \quad s_i \in \{1, \ldots, q\}, \quad (4)$$

some numerical results are compiled in Table 2 and compared with an analytical formula [35] (derived *after* the first numerical results were already published), which follows by combining (a) the assumption of complete wetting [36, 37] with (b) duality arguments [38] and (c) an exact result for the correlation length $\xi_d(\beta_t)$ in the disordered phase at the transition point $\beta_t = \ln(1 + \sqrt{q})$ [39] (see also [40, 41]),

$$2\sigma_{od} = \sigma_{oo} = 1/\xi_d = \frac{1}{4} \sum_{n=0}^{\infty} \ln\left[\frac{1 + w_n}{1 - w_n}\right], \quad (5)$$

where $w_n = \left[\sqrt{2} \cosh\left((n + \frac{1}{2})\pi^2/2v\right)\right]^{-1}$ with $v = \ln\left(\frac{1}{2}\left[\sqrt{\sqrt{q} + 2} + \sqrt{\sqrt{q} - 2}\right]\right)$. More precisely, it was shown [36] that $2\sigma_{od} \leq \sigma_{oo}$ for *all* $q \geq 5$. The opposite inequality could only be derived for $q > q_0$ (with $4 < q_0 < \infty$ being a sufficiently large constant), but by basic thermodynamic arguments it is commonly believed that it actually also holds for all $q \geq 5$.

Overall the numerical and analytical values are in good agreement, but noteworthy is the systematic trend of the numerical data to overestimate the analytical values, which are actually exact *upper* bounds.

## 3.2 Correlation Length

To test the formula (5) for the correlation length $\xi_d(\beta_t)$ directly, we performed further Monte Carlo simulations [44] of the Potts model (4) and measured the $k_y = 0$ projection $g$ of the correlation function

$$G(i, j) = \langle \delta_{s_i s_j} - 1/q \rangle, \quad (6)$$

at $\beta_t$ in the disordered phase. Here we simply choose the lattice sizes large enough in order to suppress transitions to the ordered phase. Estimates of autocorrelation times showed [44] that in this situation the optimal update procedure consists of many single-cluster [45] update steps combined with one multiple-cluster [46] update for efficient measurements using the improved estimator

$$G(i, j) = \frac{q - 1}{q} \langle \Theta(i, j) \rangle, \quad (7)$$

where $\Theta(i, j) = 1$, if $i$ and $j$ belong to the same cluster, and $\Theta = 0$ otherwise.

The numerical data for $q = 10$ at $\beta_t$ are shown in Fig. 3. From the curvature of $g(x)$ for small distances it is clear that fits with the simplest Ansatz for periodic boundary conditions, $g(x) = a \cosh((x - L/2)/\xi_d)$, will only work for very large $x$. We therefore tried to use the more general Ansatz $g(x) = a \cosh((x - L/2)/\xi_d) + b \cosh(c(x - L/2)\xi_d)$, with four parameters $a, b, c,$ and $\xi_d$. The solid line in Fig. 3 shows a fit where $\xi_d$ is held fixed at its theoretical value and only the three remaining parameters are optimized. The good quality of the fit over a wide $x$ range shows that our data are compatible with the theoretical expectations. To get a more quantitative measure we next performed four-parameter fits with $\xi_d$ as a *free* parameter. Fits over the same $x$ range yield an estimate of $\xi_d(\beta_t) = 9.2(8)$, which is about 15% smaller than the exact value but within the statistical errors still consistent. By restricting the fit interval to larger $x$ values, we observed a tendency to higher values but with the drawback of increased error bars. Since our data for $q = 15$ and 20 exhibit the same qualitative behavior, we believe that higher excitations cannot be neglected at the distances studied so far ($x_{\max} = L/2 \approx 7\xi_d$). To cope with this problem we are currently performing further simulations on $2L \times L$ lattices, which should allow to study the correlations over even larger distances.

Table 2: Comparison of analytical and numerical results for the order-disorder interface tension $2\sigma_{od}$ in 2D $q$-state Potts models.

| $q$ | $\xi_d$ | $2\sigma_{od}$ | $2\sigma_{od}$ (MC) | |
|---|---|---|---|---|
| 7 | 48.095907 | 0.020792 | 0.0241(10) | Janke *et al.* [10] |
| | | | 0.02348(38) | Rummukainen [11] |
| | | | 0.0228(24) | Grossmann and Gupta [12] |
| 8 | 23.878204 | 0.041879 | 0.045 | Janke [51] |
| 10 | 10.559519 | 0.094701 | 0.09781(75) | Berg and Neuhaus [9] |
| | | | 0.10 | Janke [51] |
| | | | 0.0950(5) | Billoire *et al.* [14] |
| 15 | 4.180954 | 0.239179 | 0.263(9) | Gupta [42] |
| 20 | 2.695502 | 0.370988 | 0.3714(13) | Billoire *et al.* [14] |

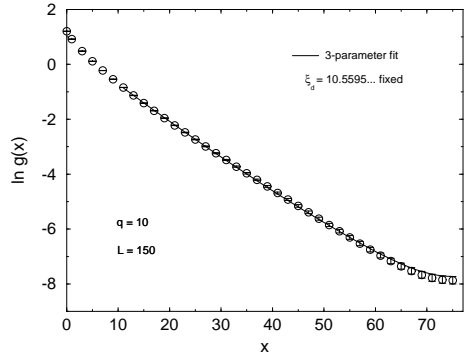

Figure 3: Semi-log plot of the correlation function $g(x)$ vs distance $x$ for $q = 10$. The solid line is a three-parameter fit with $\xi_d$ held fixed at its theoretical value.

## 3.3 Accurate Determination of the Transition Point and Latent Heat

The traditional way to locate first-order transition points is based on the scaling behavior of the specific-heat maxima $C_{\max}(V)$ and Binder-parameter minima $B_{\min}(V)$, where $C(V, \beta) = \beta^2 V(\langle e^2 \rangle - \langle e \rangle^2)$ and $B(V, \beta) = 1 - \langle e^4 \rangle / 3\langle e^2 \rangle^2$. In the idealized infinite volume limit, these quantities develop singularities at the transition point $\beta_t$. In a finite volume $V = L^d$ the singularities are smoothed out and, depending on its strength, the transition is signalized by more or less pronounced finite peaks of the specific heat or dips of the Binder parameter near the infinite volume transition point. For an illustration see Fig. 4 where data for the two-dimensional 8-state Potts model are shown. If the volume is cubic or nearly cubic the location of the extrema is typically shifted by an amount $O(V^{-1})$ with respect to the actual infinite volume transition point and one may try to estimate $\beta_t \equiv 1/T_t$ from the finite volume results by extrapolations in $1/V$ [47].

In addition to random statistical errors the data are in general also systematically shifted by exponential corrections which are difficult to take into account theoretically. This makes the extrapolation of finite volume data not always reliable. It is therefore desirable to find definitions of a finite volume transition point which involve *no* power-law corrections at all.

The starting point for such definitions is the observation that on lattices with periodic boundary conditions the partition function of a model describing the coexistence of one disordered and $q$ ordered phases can be written for large enough $q$ as [48, 49]

$$Z_{\rm per}(V, \beta) = \left( \sum_{m=0}^{q} e^{-\beta f_m(\beta) V} \right) \left( 1 + \mathcal{O}\left( V e^{-L/L_0} \right) \right). \tag{8}$$

Here $L_0 < \infty$ is a constant, $L$ is the linear length of the lattice, and $f_m(\beta)$ is the metastable free energy density of the phase $m$. It can be defined in such a way that it is equal to the idealized infinite volume free energy density $f(\beta)$ if $m$ is stable and strictly larger than $f(\beta)$ if $m$ is unstable [48, 49]. This implies (see also Refs.[50, 51]) that in the infinite

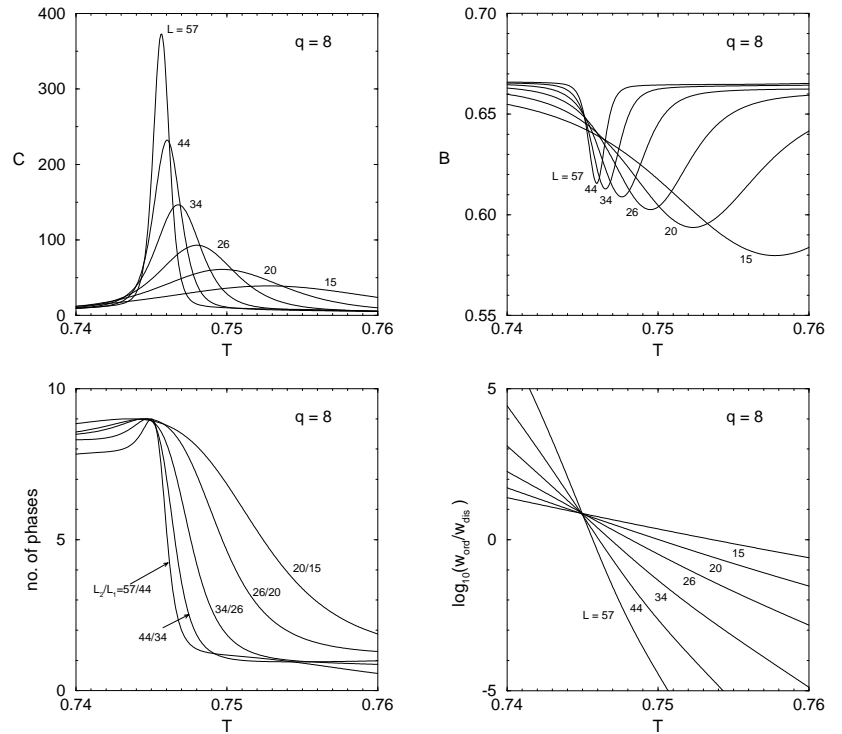

Figure 4: Finite-size scaling of the traditional observables, specific heat $C$ and Binder parameter $B$, compared with the observables defined in eqs. (10), (12) for the two-dimensional 8-state Potts model.

volume limit the parameter

$$N(V, \beta) = Z_{\rm per}(V, \beta) e^{\beta f(\beta) V} \tag{9}$$

is equal to the number of stable phases at the inverse temperature $\beta$, i.e., $N(\beta) = \lim_{V \to \infty} N(V, \beta) = q$ in the ordered phase, $N(\beta) = q + 1$ at the transition point $\beta_t$, and $N(\beta) = 1$ in the disordered phase. A natural definition of a finite volume transition point $\beta_t(V)$ is thus the point where $N(V, \beta)$ is maximal. Due to the bound (8) (and similar bounds for derivatives [48, 49]) this definition leads to only *exponentially* small shifts of $\beta_t(V)$ with respect to the infinite volume transition point $\beta_t$.

The free energy $f(\beta)$ in (9) is only defined in the thermodynamic limit and hence not accessible to numerical simulations. It is therefore necessary to eliminate this term by, e.g., forming a suitable ratio. Instead of (9) one may look for the maximum of the *number-of-phases* parameter

$$N(V_1, V_2, \beta) = \left[\frac{Z_{\rm per}(V_1,\beta)^\alpha}{Z_{\rm per}(V_2,\beta)}\right]^{\frac{1}{\alpha-1}}, \qquad (10)$$

where $\alpha = V_2/V_1$ [50, 51]. By inserting (8) it is straightforward to verify that with increasing temperature $N(V_1, V_2, \beta)$ smoothly interpolates between the values $q$, $q+1$, and 1. Actual simulation results for $\alpha = V_2/V_1 \approx 1.6$ are shown in Fig. 4. The locations of the maxima define the desired, only exponentially shifted finite-volume transition points $\beta_t(V_1, V_2)$ which will be denoted $\beta_{V/V}$. By differentiating $\ln N$ with respect to $\beta$ one readily sees that determining $\beta_{V/V}$ amounts to solving $\alpha E(V_1, \beta_{V/V}) = E(V_2, \beta_{V/V})$ or $e(V_1, \beta_{V/V}) = e(V_2, \beta_{V/V})$, i.e., to locating the crossing point of the internal energies per site, $e \equiv E/V$, of the two lattices of different size.

The numerical determination of $\beta_{V/V}$ requires simulations on two different lattices. In Ref.[50] we have therefore proposed another definition of a finite-volume transition point which requires data from one lattice only. Its definition is based on the fact that at the infinite-volume transition point all free energies $f_m(\beta)$ are equal, so that eq. (8) implies

$$w_o \equiv \sum_{m=1}^{q} e^{-\beta_t f_m(\beta_t)V} = qe^{-\beta_t f_d(\beta_t)V} \equiv qw_d, \qquad (11)$$

apart from exponentially small corrections. Here the free energy density of the "zeroth", disordered phase is denoted by $f_d$, and $w_{o,d}$ are the associated statistical weights of the coexisting phases. A natural definition of a finite-volume transition point $\beta_W$ is thus the point where the ratio of the total weight of the $q$ ordered phases to the weight of the disordered phase approaches $q$. More precisely we introduce the *ratio-of-weights* parameter

$$R(V, \beta) \equiv W_o/W_d \equiv \sum_{E<E_0} P_{V,\beta}(E) / \sum_{E \geq E_0} P_{V,\beta}(E), \qquad (12)$$

where $P_{V,\beta}(E)$ are the (double-peaked) energy histograms, and determine $\beta_W$ by solving

$$R(V, \beta_W) = q. \qquad (13)$$

The parameter $E_0$ in (12) is defined by reweighting [52] the energy distribution to the temperature where the two peaks of $P_{V,\beta}(E)$ have equal height and then taking $E_0$ as the energy at the minimum between the two peaks. Other definitions of $E_0$ would be reasonable as well, as for example the internal energy at the temperature where the specific heat is maximal. Since it is expected that the relative height of the minimum between the two peaks decreases like $\exp(-2\sigma L^{d-1})$ as $L \to \infty$, all these definitions do in fact only differ by exponentially small corrections and it is a matter of practical convenience to choose $E_0$. In Fig. 4 the logarithm of the ratio-of-weights parameter $R(V, \beta)$ is plotted as a function of temperature.

In (13) we have assumed that the number of ordered phases, $q$, is known by general arguments. If this is not the case, one may use the crossing points $\beta_{W/W}$ satisfying $R(V_1, \beta_{W/W}) = R(V_2, \beta_{W/W})$ as estimates for $\beta_0$. The value of $R$ at the crossing point then gives the ratio of the number of coexisting ordered and disordered phases. This, however, requires again the simulation on two lattices of different size.

11

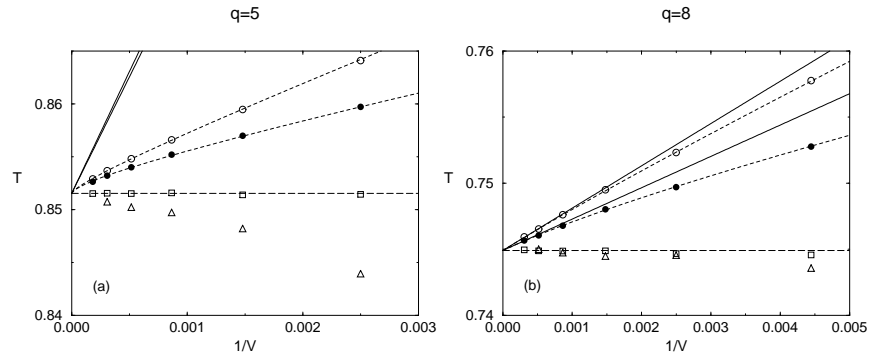

Figure 5: Scaling behavior of the finite-volume transition points defined by the Binder-parameter minimum ($\circ$), the specific-heat maximum ($\bullet$), the maximum of the number-of-phases parameter ($\triangle$), and the ratio-of-weight parameter ($\square$). The solid straight lines are the exactly known $1/V$ corrections corresponding to $\circ$ and $\bullet$, and the dashed, almost interpolating curves show exponential fits (including the $1/V$ corrections) to these data. Note that the $(1/V)^2$ corrections are almost invisible on this scale and in any case point in the "wrong" upward direction. The long dashed horizontal lines indicate the exact infinite-volume transition points.

The convergence properties of the four finite-volume transition points are compared in Fig. 5 for a very weak ($q = 5$) and a rather strong ($q = 8$) first-order transition. Notice the surprisingly fast convergence of $\beta_W(V)$ even for $q = 5$.

The ratio-of-weights method leads naturally to a finite-volume definition of the latent heat [51] which also should have only exponentially small corrections with respect to the infinite volume limit. Since

$$\ln(w_o/w_d) = -\beta V(f_o - f_d), \qquad (14)$$

the slopes of $R(V, \beta)$ in Fig. 4 at the crossing point may be used to define

$$\Delta e(V) = e_d(V) - e_o(V) = \frac{d}{d\beta}\ln(W_o/W_d)/V = -\frac{1}{T^2}\frac{d}{dT}\ln(W_o/W_d)/V. \qquad (15)$$

The resulting estimates $\Delta e(V)$ are plotted in Fig. 6 and compared with the traditional definition based on the peak locations of $P_{V,\beta}(E)$ [53]. For strong first-order transitions ($q = 8$ and 10) the asymptotic limit is indeed reached much faster with the new definition. For a very weak transition ($q = 5$), on the other hand, both methods yield comparable estimates which are still far away from the limiting value.

12

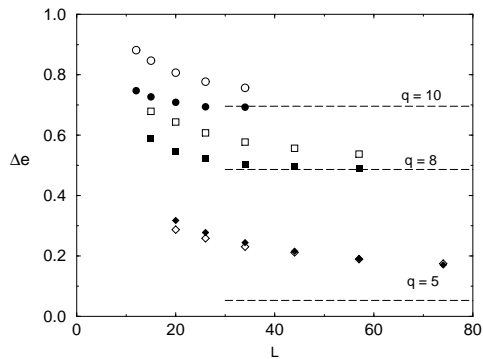

Figure 6: The finite-volume latent heat $\Delta e$ of the 2D $q$-state Potts model vs linear lattice size $L$. The open symbols show the traditional estimates from the peak locations of $P_{V,\beta}(E)$, and the filled symbols follow from the slopes of the ratio-of-weight parameter. The dashed horizontal lines show the exactly known infinite-volume limits [54].

## 4 Concluding Remarks

The material reviewed in these lecture notes is, admittedly, a somewhat biased selection of recent developments in this field. Different improved Monte Carlo update schemes are discussed, e.g., in Refs.[55, 56]. Also, refined data analyses have been developed along many different lines. For a recent review focusing on the analysis of probability distributions and various cumulants, e.g., see Ref.[57].

The combination of multicanonical reweighting with multigrid update schemes leads to a significant reduction of autocorrelation times by a roughly constant factor. It still remains a challenge to develop Monte Carlo algorithms which exhibit no slowing down at a first-order phase transition.

## Acknowledgments

I would like to thank Bernd Berg, Christian Borgs, Stefan Kappler, Mohammad Katoot and Tilman Sauer for fruitful collaborations and the DFG for a Heisenberg fellowship.